\documentclass[11pt]{article}
\usepackage{blois,epsfig}
\usepackage[latin1]{inputenc}
\usepackage[OT1]{fontenc}
\usepackage{rotating}

\def\be{\begin{equation}}
\def\ee{\end{equation}}
\def\bea{\begin{eqnarray}}
\def\eea{\end{eqnarray}}

\newcommand{\lsim}   {\mathrel{\mathop{\kern 0pt \rlap
  {\raise.2ex\hbox{$<$}}}
  \lower.9ex\hbox{\kern-.190em $\sim$}}}
\newcommand{\gsim}   {\mathrel{\mathop{\kern 0pt \rlap
  {\raise.2ex\hbox{$>$}}}
  \lower.9ex\hbox{\kern-.190em $\sim$}}}
\def\be{\begin{equation}}
\def\ee{\end{equation}}
\def\ba{\begin{eqnarray}}
\def\ea{\end{eqnarray}}

\begin{document}
\vspace*{0.1cm}
\title{THE RISE AND FALL OF TOP-DOWN MODELS\\[0.5ex] AS MAIN UHECR SOURCES}

\author{M.~Kachelrie\ss}

\address{Institutt for fysikk, NTNU Trondheim, Norway}

\maketitle

\abstracts{
The motivation and the current status of top-down models as sources of  
ultrahigh energy cosmic rays (UHECR) are reviewed. Stimulated by the AGASA 
excess, 
they were proposed as the main source of UHECRs beyond the GZK cutoff. 
Meanwhile searches for their signatures have limited their contribution 
to the UHECR flux to be subdominant, while the theoretical motivation 
for these searches remained strong: Topological defects are a generic 
consequence of Grand Unified Theories and superheavy particles are a 
creditable dark matter candidate.  While Fermi/GLAST results should help to
improve soon bounds on topological defects from the  diffuse 
gamma-ray background, the most promising detection method are UHE neutrino 
searches. Superheavy dark matter can be restricted or detected by its 
characteristic galactic anisotropy combined with searches for UHE 
photons. 
}

\section{Introduction}

Cosmic Rays (CR) are observed in an energy range extending over more than 
eleven decades, with a flux described by a nearly featureless power-law 
starting from subGeV energies up to $3\times 10^{20}$~eV. 
Soon after the first CR with energy beyond $10^{20}\,$eV was observed in 1963
at the Vulcano Ranch experiment, it was shown that the universe is filled
with thermal relic photons from the big-bang.
Greisen\,\cite{G} and independently Zatsepin and Kuzmin\,\cite{ZK} 
pointed out that as a consequence of pion-production on these photons, 
$p+\gamma_{3K}\to\Delta^\ast\to N+\pi$, the energy-loss-length 
$(1/E)({\rm d}E/{\rm d}t)$ of protons should decrease dramatically around 
$E_{\rm GZK}\sim 5\times 10^{19}\,$eV.
Nuclei exhibit an even more pronounced cutoff at a (for iron) somewhat higher
energy, while photons are absorbed over a few Mpc due to pair-production
on the radio background. Thus, the ultrahigh energy (UHE) CR spectrum should 
steepen at $E_{\rm GZK}$ for {\em any\/} homogeneous distribution of proton 
or nuclei sources\,\cite{prop}.  

It is the decrease of the UHECR horizon scale below $\sim 100$\,Mpc 
that may make astronomy with charged particles possible: Within this 
distance, the positions of powerful objects like active galactic nuclei 
are known and the effect of deflections of protons in (extra-) galactic 
magnetic fields\,\cite{egmf} may be sufficiently small to allow us the 
identification of sources. Even if the identification of individual 
sources may be not possible, the
large-scale structure of sources is not yet averaged out along the
line-of-sight and results in specific anisotropy patterns\,\cite{ani,aniso}.
The ``GZK puzzle'' of the 90s was the observation of a 
surprisingly large number of events above $10^{20}\,$eV, in particular by
the AGASA  experiment, while no promising nearby sources could be identified in 
the direction of these UHECRs. Moreover, the arrival directions of UHECRs 
were found to be consistent with isotropy on medium and large angular scales
prior to an analysis\,\cite{ani} of the cumulative auto-correlation function
of all publicly available data, while the significance of the observed
small-scale clustering\,\cite{ssc} remained disputed.
Finally, the acceleration of CRs to energies beyond $10^{20}\,$eV was
(and still is) a challenge for all known astrophysical sources.

The extension of the UHECR spectrum beyond the  GZK cutoff as observed by  
AGASA, the missing correlation of the UHECR arrival directions with powerful 
nearby sources and the difficulty to accelerate particles in astrophysical 
accelerators up to energies $E\gsim 10^{20}$~eV prompted models that do not 
accelerate charged particles (``bottom-up models'') but use the stable 
secondaries produced in the fragmentation of superheavy particles 
(``top-down models''). Note that the study of both
main variants of top-down models, superheavy dark matter (SHDM) and 
topological defects (TD), is well motivated independently of UHECR
observations: The latter are a generic prediction of Grand Unified 
Theories, in particular if supersymmetry is included\,\cite{susygut}, while
brane inflation\,\cite{brane} would result in the production of macroscopic 
superstrings. Superheavy dark matter is an attractive DM candidate, because
(meta-) stable particles with masses around $10^{13}$\,GeV are produced 
during inflation with the correct abundance\,\cite{grav} to be today the 
main source of DM. In the following, we up-date the 
review\,\cite{Kachelriess:2004ax} of the status of these 
models.

\section{Top--down models and their signatures}
Top--down model is a generic name for all proposals in which the
observed UHECR primaries are produced as decay products of some
superheavy particles $X$. These $X$ particles can be either metastable
or be emitted by topological defects at the present epoch. In both
cases, the range of masses that was suggested by the AGASA 
excess\,\cite{AGASA} is rather narrow, 
$10^{12}\,{\rm GeV}\lsim m_X\lsim {\rm few}\times  10^{13}$~GeV. The lower 
limit follows directly from the highest CR energies observed, 
$E_{\max}=(2-3)\times 10^{20}$\,eV, while the upper limit can be derived by
comparing the integral flux predicted in these models with the 
non-observation of UHECRs above $E_{\max}$.

\subsection{Superheavy dark matter}
The suggestion\,\cite{bkv97,kr97} of superheavy metastable relic particles 
as UHECR sources was originally motivated by the AGASA excess, although it 
was soon realized that stable or metastable particles in this mass range 
are generically good DM candidates\,\cite{grav}. Since they constitute
(part of) the CDM, their abundance in the galactic halo is enhanced by a 
factor $\sim 2\times 10^5$ above their extragalactic abundance. 
Therefore, the proton and photon flux is dominated by the halo
component and the GZK cutoff is avoided\,\cite{bkv97}. 
The ratio $r_X=\Omega_X (t_0/\tau_X)$ of the product of the relic abundance 
$\Omega_X$ and the age of the universe $t_0$ and the lifetime $\tau_X$ 
of the $X$ particle is  $r_X\sim 10^{-11}$, if the UHECR flux is fixed to 
the observations of 
AGASA. The numerical value of $r_X$ is not predicted in the generic 
SHDM model, but calculable as soon as a specific particle physics and 
cosmological model is fixed.


Various mechanisms exist that produce particles with a non-thermal distribution
in the early Universe. Since the energy density of non-relativistic particles 
decreases slower than the one of radiation, their abundance increases by the 
factor $a(t_0)/a(t_\ast)$ with respect to radiation, where $a(t_0)$ and 
$a(t_\ast)$ are the scale factors of the universe today and at the epoch of 
particle generation, respectively. If particle
production happens  at the earliest relevant time, i.e.\ during inflation, this 
factor can become extremely large, $\sim 10^{22}$. Not surprisingly, 
such a small energy fraction can be transferred to SHDM by many different 
mechanisms,  as thermal production at reheating\,\cite{bkv97,ChKR-r}, the   
non-perturbative regime of a broad parametric resonance at 
preheating\,\cite{preheating}, and production by topological 
defects\,\cite{bkv97,Kolb}.

Here we recall only the generation of SHDM by gravitational interactions 
from vacuum fluctuations at the end of inflation\,\cite{grav,grav2}.
Numerical calculations\,\cite{grav2} for the present abundance 
of fermionic SHDM can be approximated as
\be \label{abund}
 \Omega_X h^2 \approx \frac{T_R}{10^{8}\,{\rm GeV}} 
           \:\left\{
           \begin{array}{ll} 
            \left( m_X/H_I \right)^2,
            & \qquad m_X\ll H_I \\
            \exp( -m_X/H_I)  \,,     
            & \qquad m_X\gg H_I 
           \end{array}
           \right.
\ee
with $H_I\approx 10^{13}$\,GeV as value for the Hubble parameter during 
inflation. 
In general, two values of the SHDM mass are consistent with the 
observed abundance $\Omega_{\rm CDM}=0.105$ of CDM\,\cite{WMAP5} for a 
specific value of the reheating temperature $T_R$. 
Choosing the larger of the two possible masses in Eq.~(\ref{abund})
together with the highest value allowed by the gravitino problem for
the reheating temperature, $T_R=10^9$\,GeV, leads to an abundance of
SHDM as observed by WMAP, while secondaries of SHDM decays 
could explain at the same time (part of) the cosmic rays at the highest 
energies, $E\gsim 10^{20}$\,eV.

The lifetime of the superheavy particle has to be in the range
$t_0\sim 10^{17}~{\rm s}\lsim \tau_X\lsim 10^{28}$~s, i.e. longer or much
longer than the age of the Universe, if UHECRs are produced by SHDM decays. 
Therefore it is an obvious
question to ask if such an extremely small decay rate can be obtained
in a natural way. A well-known example of how metastability can be
achieved is the proton: In the standard model B--L is a conserved
global symmetry, 
and the proton can decay only via non-renormalizable operators. 
Similarly, the $X$ particle could be protected by a new global symmetry
which is only broken by higher-dimensional operators suppressed by
$M^{d}$, where for instance $M\sim M_{\rm Pl}$ and $d\geq 7$ is possible. 
The case of discrete gauged symmetries has been studied in detail\,\cite{Ha98}. Another possibility is that the global symmetry is
broken only non-perturbatively, either by wormhole\,\cite{bkv97} or
instanton\,\cite{kr97} effects. Then an exponential suppression of the
decay process is expected and lifetimes $\tau_X\gsim t_0$ can be
naturally achieved. 

An example of a metastable SHDM particle in a semi-realistic
particle physics model is the crypton\,\cite{El90}.
Cryptons are bound-states from a strongly interacting hidden sector of
string/M theory. Their mass is determined by the non-perturbative
dynamics of this sector and, typically, they decay only through
high-dimensional operators. For instance, flipped SU(5) motivated by
string theory contains bound-states with mass $\sim 10^{12}$~GeV and 
$\tau\sim 10^{15}$~yr\,\cite{Be99}. Choosing $T_R\sim 10^5$~GeV results
in $r_X\sim 10^{-11}$, i.e. the required value to explain the UHECR
flux above the GZK cutoff. This example shows clearly that the SHDM
model has no generic ``fine-tuning problem.'' Other viable candidates
suggested by string theory were also discussed\,\cite{cfp}.
Another well-suited candidate for superheavy dark matter is the lightest 
supersymmetric particle within the scenario of superheavy supersymmetry\,\cite{SS}.
This is a unique case where the SHDM has weak interactions and respects
perturbative unitarity despite the large mass hierarchy $m_X\gg m_Z$.

Finally, we comment on the case of stable SHDM. Since the annihilation
cross section of a (point) particle is bounded by unitarity, 
$\sigma_{\rm ann}\propto 1/M_X^2$, the resulting flux of UHE secondaries is 
too small to be observable without an additional enhancement of the 
annihilation rate.

\subsection{Topological defects}
Topological defects\,\cite{td}  such as (superconducting)
cosmic strings,  monopoles, and hybrid defects can be
effectively produced in non-thermal phase transitions during the
preheating stage\,\cite{Kh98/Ku99}. Therefore the presence of TDs
is not in conflict with an inflationary period of the early
Universe. They can naturally produce particles with high enough
energies but have problems to produce large enough fluxes of UHE
primaries, because of the typically large distance between TDs. Then the
flux of UHE particles is either exponentially suppressed or strongly
anisotropic if a TD is nearby by chance.

{\em Ordinary strings\/} can produce UHE particles, e.g., when string loops
self-intersect or when two cusp segments overlap and annihilate. In
the latter case, the maximal energy of the produced fragmentation
products is not $m_X/2$, but can be much larger due to the high
Lorentz factors of the ejected $X$ particles.

{\em Superconducting strings:\/}
Cosmic strings can be superconducting in a broad class
of particle models. Electric currents can be induced in
the string either by a primordial magnetic field that decreases during
the expansion of the Universe or when the string moves through
galactic fields at present. If the current reaches the vacuum
expectation value of the
Higgs field breaking the extra U(1), the trapped particles are ejected
and can decay.

{\em Monopolium M,\/} a bound-state of a monopole--antimonopole pair, was
the first TD proposed as UHECR source\,\cite{Hi83}. It clusters like
CDM and is therefore an example for SHDM. The
galactic density of monopoles is constrained by the Parker limit: the
galactic magnetic field should not be eliminated by the acceleration of
monopoles. Reference\,\cite{Bl99} concluded that the resulting limit on
the UHECR flux produced by  Monopolium annihilations is 10 orders of
magnitude too low.

{\em Cosmic necklaces\/} are hybrid defects
consisting of monopoles connected by a string. These defects are produced 
by the symmetry breaking $G\to H\times U(1) \to H\times Z_2$, where
$G$ is semi-simple.
In the first phase transition at scale $\eta_m$, monopoles are 
produced. At the second phase transition, at scale  $\eta_s<\eta_m$, each 
monopole gets attached to two strings. 
The basic parameter for the evolution of necklaces is the ratio
$r=m/(\mu d)$ of the monopole mass $m$ and the mass of the string 
between two monopoles, $\mu d$, where $\mu \sim \eta_s^2$ is the mass 
density of the string and $d$ the distance between two monopoles.
While for $r\ll 1$ necklaces evolve in the scaling regime of a usual
string network, monopoles dominate the evolution in the opposite limit
$r\gg 1$. Analytical estimates\,\cite{Be97} suggested that for a reasonable 
range of
parameters strings lose their energy and contract through gravitational 
radiation in such a way that a sufficient fraction of monopole-antimonopole
pairs annihilate in the nearby universe at the present epoch producing 
$X$ particles. In particular, it was argued\,\cite{Be97,Aloisio:2003xj} 
that the model predicts a UHECR flux close to the observed one without 
violating the EGRET bound. Numerical 
studies\,\cite{Bl99,Si00,BlancoPillado:2007zr} 
of the evolution of necklaces
found that the lifetime of necklaces is generally much shorter than
the age of the Universe. 
More recently, it has been argued\,\cite{BlancoPillado:2007zr} 
that monopoles attached to the strings acquire large velocities,
leading to fast annihilations or to the escape of monopole-antimonopole 
pairs. In both cases, necklaces would have been transformed to a standard
cosmic string network at the present epoch.

\subsection{Signatures of top-down models}
Superheavy dark matter  has several clear signatures: 
1. No GZK cutoff, instead a flatter spectrum compared to astrophysical
sources up to the kinematical cutoff close to $m_X/2$. 
2. Large neutrino and photon fluxes compared to the proton flux.
3. Galactic anisotropy. 
4. If low-scale supersymmetry is realized by nature and $R$ parity conserved, 
then the lightest supersymmetric particle (LSP) is an additional UHE 
primary. 
5. No correlation of the CR arrival directions with astrophysical sources 
at the highest energies, $E\gsim 10^{20}$\,eV.

\begin{figure}[ht]
\includegraphics[width=0.52\textwidth]{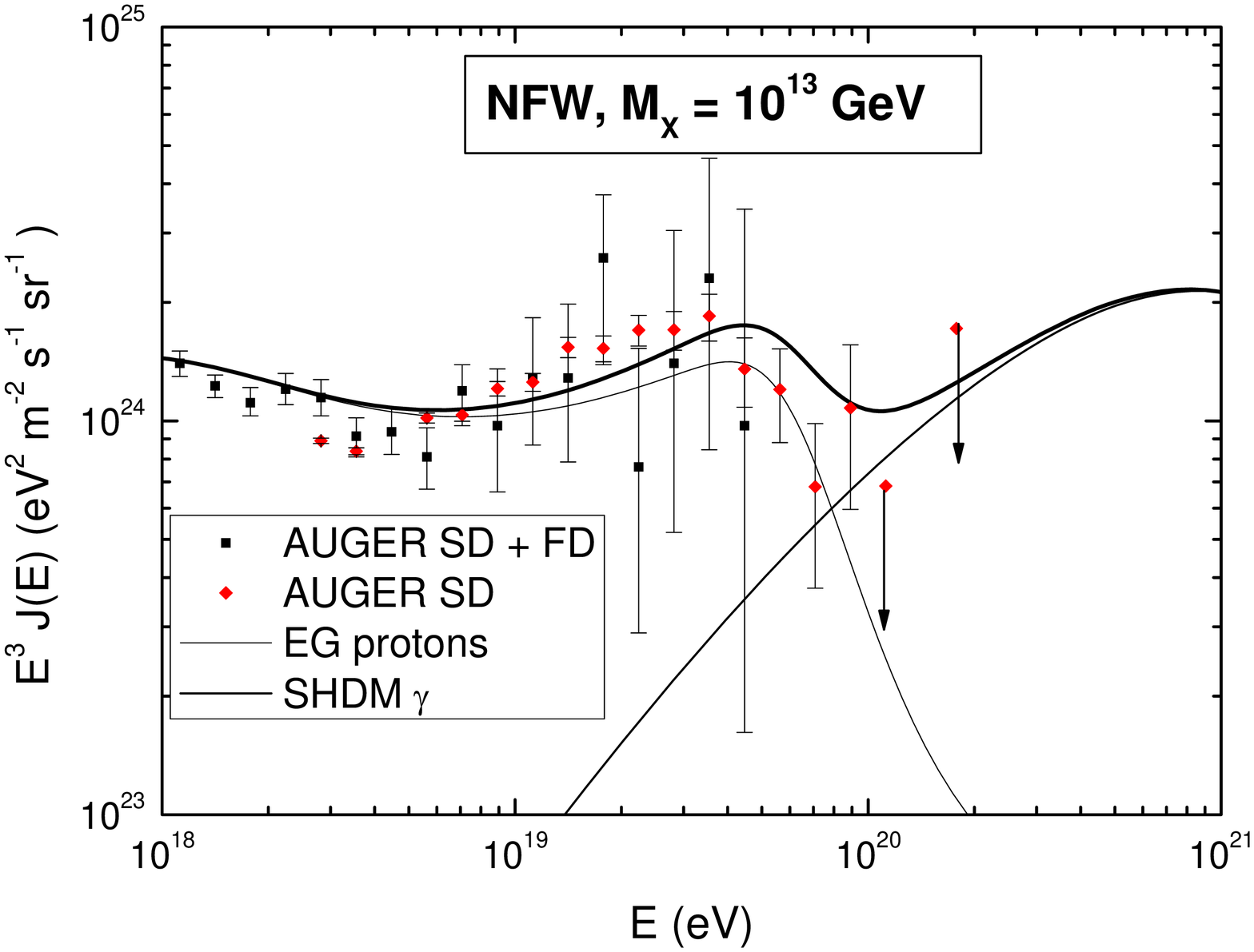}
\includegraphics[width=0.52\textwidth]{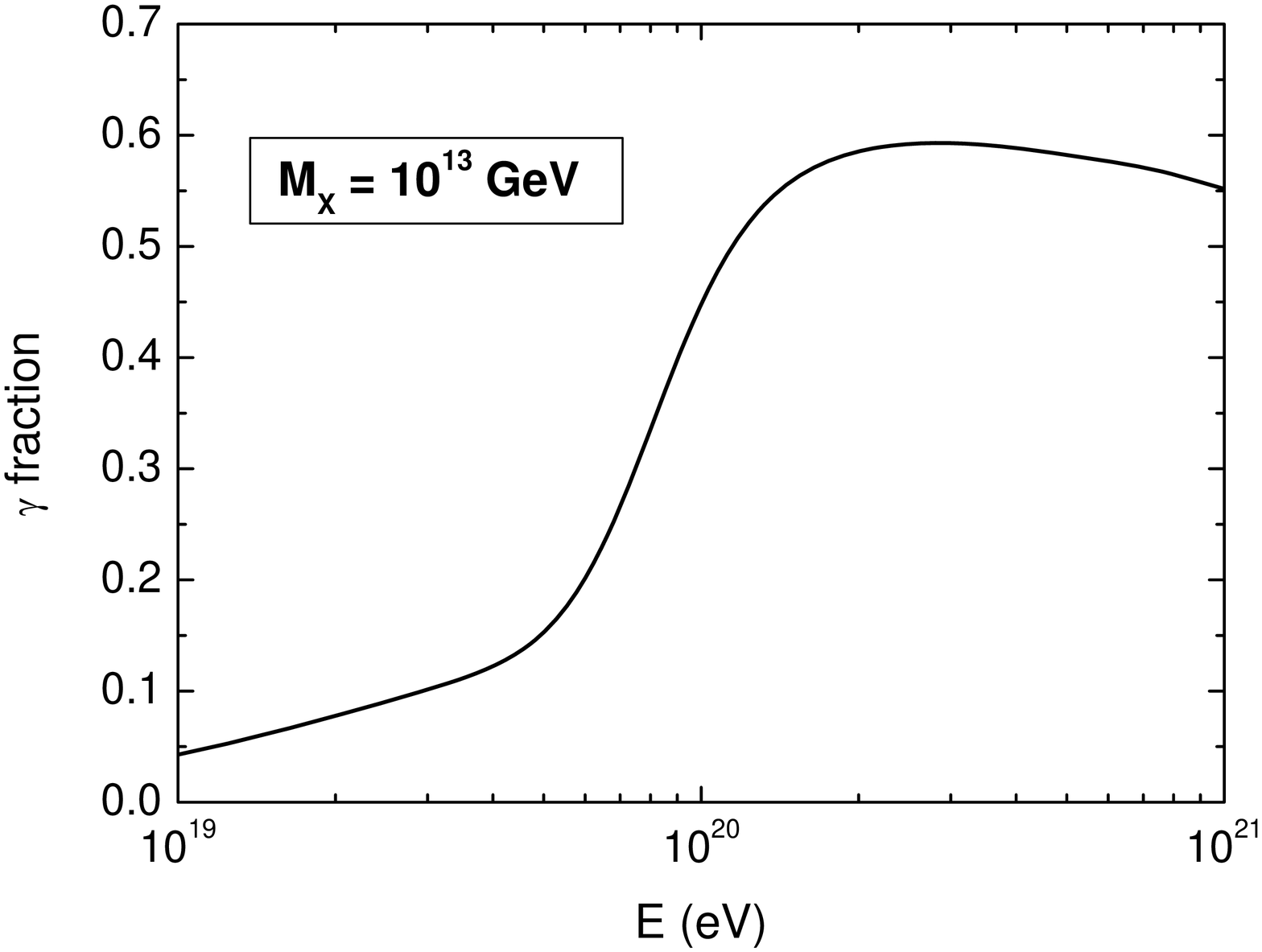}
\caption{\label{ABK}
Left: The calculated spectrum of UHECRs from SHDM (dotted curve) in 
comparison with the Auger data together with the spectrum of extragalactic 
protons. 
Right: Fraction of galactic SHDM photons in the total UHECR flux for the Auger 
detector assuming a NFW density profile; both figures from Ref.~[57].}
\end{figure}

{\em 1. Spectral shape:\/}
The fragmentation spectra of superheavy particles calculated by
different methods and different groups~\cite{Be00,tb,Aloisio:2003xj}
agree quite well, cf. Ref.~\cite{Aloisio:2003xj} for a detailed
discussion. This allows one to consider the spectral shape as a signature
of models with decays or annihilations of superheavy particles. The 
predicted spectrum in the SHDM model, $dN/dE\propto E^{-1.9}$, cannot 
fit the observed UHECR spectrum at energies $E\leq (6$--$8)\times
10^{19}$~eV. Thus only events at $E\geq (6$--$8)\times 10^{19}$~eV, and in
particular the AGASA excess could be explained in this model.
Discarding the AGASA results, the necessity for SHDM decays
is obviously reduced: A two-component fit\,\cite{Aloisio:2007bh} to the 
experimental data from the PAO using protons from uniformly, continuously 
distributed extragalactic  astrophysical sources and photons from SHDM 
is shown in  the left panel of Fig.~\ref{ABK}. Although the quality of the 
fit including SHDM is good, it is clear that the presence of the cutoff
in the energy spectrum makes an additional component at $E>10^{20}$eV
unnecessary.

The UHECR flux in topological defects like cosmic strings has a GZK cutoff 
that is however less pronounced than for astrophysical sources, because of 
the flatter generation spectrum of the UHE particles and the dominance of
photons. The resulting shape made it, even neglecting all other 
constraints, impossible to explain the AGASA excess with TDs.

{\em 2. Chemical composition:\/}
Since at the end of the QCD cascade quarks combine more easily to
mesons than to baryons, the main component of the UHE flux are
neutrinos and photons from pion decay.
Therefore, a robust prediction of this model is photon dominance with
a photon/nucleon ratio\,\cite{Aloisio:2003xj} of $\gamma/N \simeq 2$--3, 
becoming smaller at the largest $x=2E/M_X$. This ratio can be
reduced for TD models because of the strong absorption of UHE photons, 
but is still much higher than expected from astrophysical 
sources\,\cite{Aloisio:2003xj,dima}.
Large UHE neutrino fluxes, especially of TD models, are another signature
of top-down models. 

Since the mass scale $M_X$ is far above those tested at accelerators, one
may wonder how new physics between the weak scale and $M_X$ can influence 
the $\nu:\gamma:N$ ratio. The answer is that this ratio is extremely stable, 
since the underlying $\pi^\pm:\pi^0:N$ ratio at hadronization, 
$Q^2\sim \Lambda_{\rm QCD}^2$, is only weakly dependent
on the ``prehistory'' of the QCD cascade at higher virtualities. One may
consider the small changes induced by weak-scale supersymmetry in the 
secondary spectra as one example\,\cite{Be00}, while a more extreme
example is a SHDM particle that is coupled at tree-level only to 
neutrinos\,\cite{Gelmini:1999ds}: Even in this case, the resulting 
electroweak cascade leads to a large fraction of photons as 
secondaries\,\cite{Berezinsky:2002hq}.

{\em 3. Galactic anisotropy:\/} 
The UHECR flux from SHDM should show a galactic
anisotropy\cite{Du98}, because the Sun is not in the center of the
Galaxy. The magnitude of this anisotropy depends on how strong the CDM is
concentrated near the galactic center. Since the decay rate is in
contrast to annihilation rate only linear in the number density $n_X(r)$, 
differences in the CDM profile impact not too much the resulting anisotropy.

{\em 4. LSP as UHE primary:\/}
An experimentally challenging but theoretically very clean signal both
for supersymmetry and for top-down models would be the detection of
the LSP as an UHE primary~\cite{Berezinsky:1997sb,Berezinsky:yk}.   
A decaying supermassive $X$ particle initiates a particle cascade
consisting mainly of gluons and light quarks but also of gluinos,
squarks and even only electroweakly interacting particles for
virtualities $Q^2\gg m_W^2, M_{\rm SUSY}^2$. When $Q^2$ reaches
$M_{\rm SUSY}^2$, the probability for further branching of
the supersymmetric particles goes to zero and their decays
produce eventually UHE LSPs. Signatures of UHE LPSs are 
a Glashow-like resonance at $10^9$~GeV $M_e/$TeV, where $M_e$ is the
selectron mass, and up-going showers for energies where the Earth is
opaque to neutrinos~\cite{Berezinsky:1997sb,BDHH}.

{\em 5. Correlations:\/}
Since in the SHDM model the contribution from other galaxies is strongly
suppressed and other TDs like cosmic strings do not cluster, no correlation
of the CR arrival directions with matter is expected at the highest energies, 
$E\gsim 10^{20}$\,eV. Hence already the correlation of UHECR arrival 
directions with matter along the supergalactic plane suggested by the
PAO data excludes top-down models in the relevant energy range,
while a correlation with a specific astrophysical source class like
active galactic nuclei would be a conclusive proof for the dominance
of astrophysical sources in the considered energy range.

\section{Observational constraints}

\subsection{UHECR observations}

\begin{figure*}[ht]
\begin{center}
   \begin{minipage}[ht]{50mm}
     \centering
     \includegraphics[width=50mm]{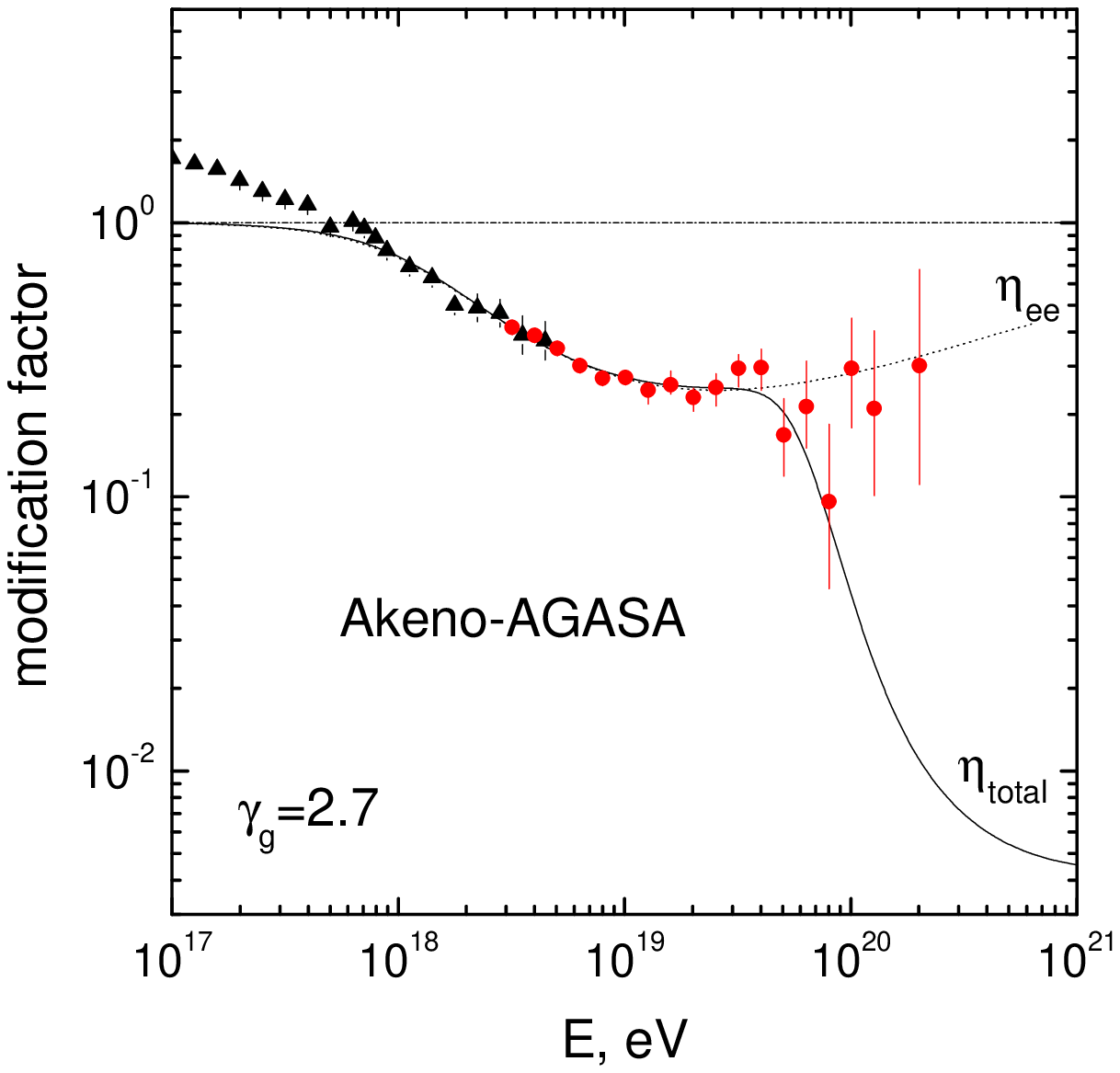}
   \end{minipage}
   \hspace{1mm}
   \vspace{-1mm}
   \begin{minipage}[ht]{51mm}
     \centering
     \includegraphics[width=50 mm]{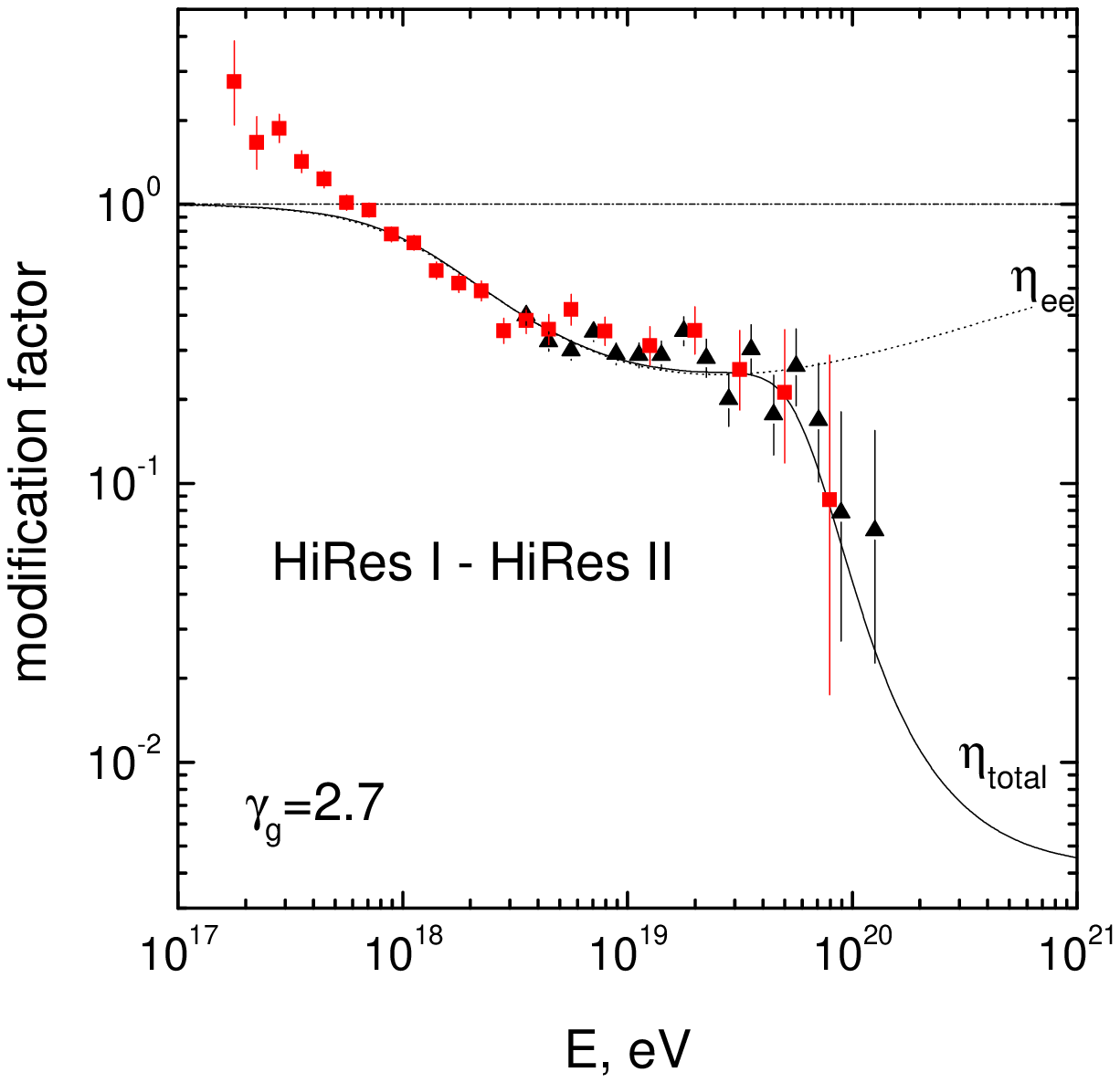}
   \end{minipage}
   \hspace{1mm}
   \vspace{-1mm}
 \begin{minipage}[h]{50mm}
    \centering
    \includegraphics[width=50 mm]{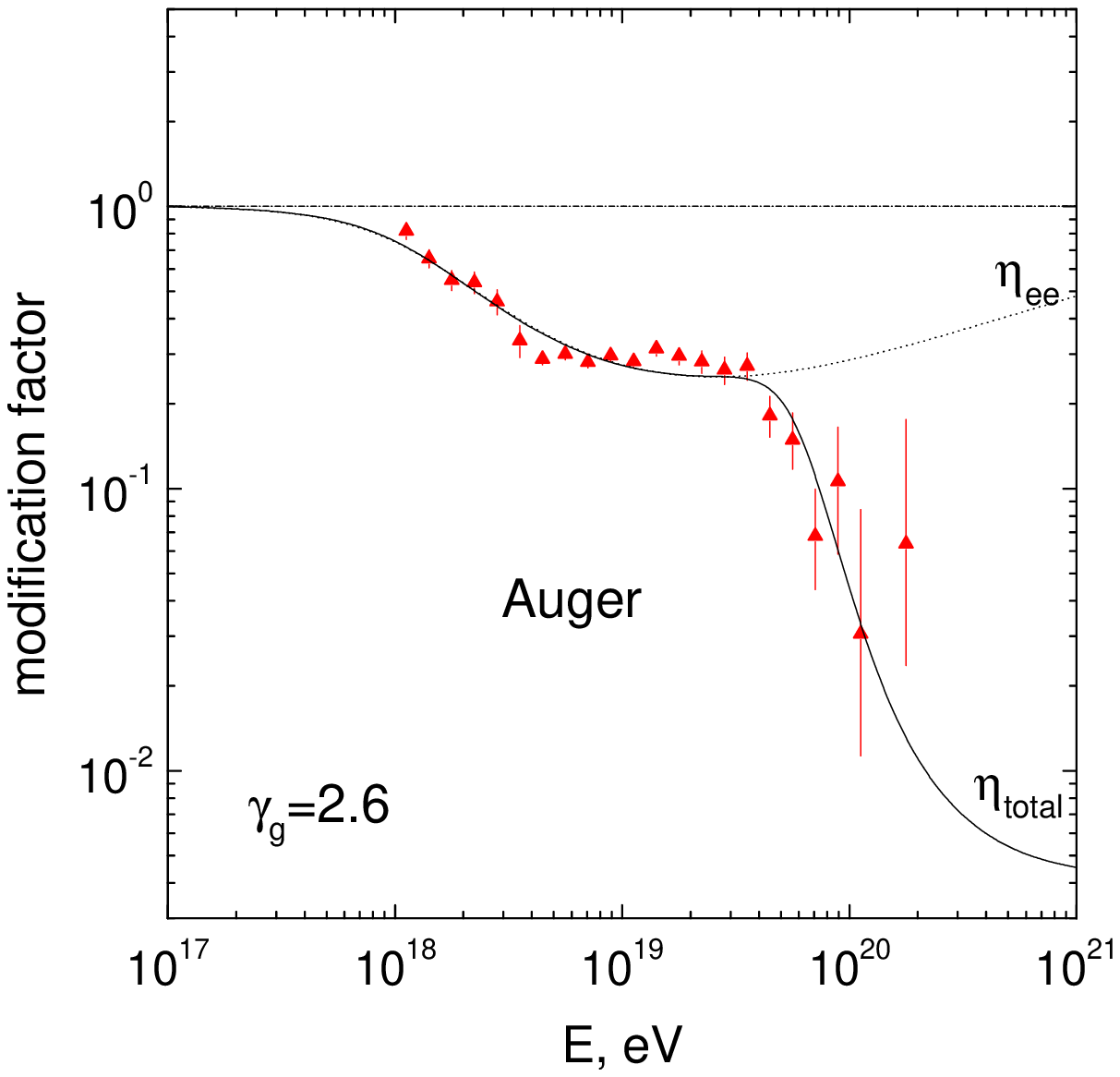}
 \end{minipage}
\end{center}
%
\caption{ The pair-production dip and GZK suppression as  predicted by
simplest model of uniformly distributed proton sources compared to
  Akeno-AGASA, HiRes, and Auger data, from Ref.~[45].
\label{fig:dip}} 
\end{figure*}

{\em 1. Spectral shape and the GZK suppression:\/}
Figure~\ref{fig:dip} shows the modification factor,\cite{BG88} 
$\eta(E)=J_p(E)/ J_p^{\rm unm}(E)$, i.e.\ the ratio of the spectrum $J_p(E)$ 
calculated with all energy losses and of the unmodified spectrum 
$J_p^{\rm unm}(E)$ calculated with adiabatic energy losses only,
for the simplest model of uniformly distributed proton sources 
together with data\,\cite{data} from Akeno-AGASA, HiRes, and PAO.
The predicted modification factor $\eta(E)$ shows a dip due to
pair-production ($\eta_{e}$) and a subsequent suppression by roughly
two orders of magnitude due to the GZK effect ($\eta_{\rm total}$).
After rescaling their energies within the range  allowed by the
experimental uncertainties, the data of all three experiments show
good agreement with the prediction in the dip region. In the energy
range where the flux is mainly modulated by the GZK effect, the
two experiments with the largest exposure (PAO and HiRes) follow too the flux
expected from uniformly distributed astrophysical sources, while
the data of the AGASA experiment show an excess above 
$10^{20}$\,eV. 
The prevailing interpretation of the AGASA excess has changed with time:
Nowadays it is considered as an experimental artefact, although its exact 
origin is unclear. Additionally to the problem of small-number statistics,
cosmic variance and potential problems with the energy calibration one should
note that also the PAO data show a certain tension between the energy 
determination using surface and fluorescence detectors.

Experimentalists prefer to quantify the significance of the GZK suppression
in a model-independent way by fitting a broken power-law 
$J\propto  E^{-\gamma_i}$ to their data above  $\sim 10^{19}\,$eV.
The HiRes collaboration\cite{Abbasi:2007sv} was the first to find with
this method a $5\sigma$ evidence for the GZK suppression: The best-fit
power-law to their data has an exponent $\gamma=5.1$ above the break energy 
$5.6\times 10^{19}\,$eV and $\gamma=2.8$ below the break down to the ankle at 
$E=4.5\times 10^{18}\,$eV.
Extrapolating the $\gamma=2.8$ spectrum above $5.6\times 10^{19}\,$eV, 
calculating the expected event number above the break and comparing with the 
observed one, the HiRes collaboration derived a 5.3\,sigma deficit.
The PAO collaboration~\cite{Collaboration:2008ru} presented more recently
a similar analysis. 
Using a method which is independent of the true slope the CR spectrum, they 
rejected a straight power-law extension above $4\times 10^{19}$\,eV with more 
than  $6\sigma$.
In both data sets, the energy where the flux is reduced by a factor two
coincides reasonably well with theoretical calculations.

In summary, the spectral shape of the observed spectra of HiRes, PAO 
and older experiments like Yakutsk and Fly's Eye are in good agreement 
with the simplest model of uniformly
distributed astrophysical sources of protons. The AGASA excess is
interpreted as an experimental artefact. As a result the SHDM flux
is not fixed, but bounded from above by the experimental data.

\noindent
{\em 2. Chemical composition:\/}
Extensive air showers induced by photons can be distinguished by
several properties from showers initiated by protons or nuclei:
First, the muon content of photon induced EAS is smaller compared
to hadronic showers. Second, photon showers have a large elongation rate
$dX_{\max}/dE$ and develop thus their maxima deeper 
in the atmosphere. Finally, the geomagnetic effect induces an 
characteristic anisotropy in the arrival directions of photon showers.

The muon content of air showers was used first by the AGASA 
experiments to derive an upper limit on the fraction of photons
in UHECRs~\cite{AGASA-gamma}. 
From eleven events at $E>1\times 10^{20}$~eV, six were measured 
within in a sub-array equipped with muon detectors. In two of them with 
energies about $1\times 10^{20}$~eV, the muon density was almost twice
higher than predicted for gamma-induced EAS, while  the muon content of the
remaining four EAS only marginally agreed with that predicted for
gamma-induced showers.  
Improved limits were later derived~\cite{Rubtsov:2006tt} by a combination of 
AGASA and Yakutsk data using the same technique.

The PAO collaboration derived photon limits using $X_{\max}$ measurement
from the fluorescence method~\cite{l1} and using a hybrid approach~\cite{l2}. 
No photon-like events were reported and the resulting limits on the fraction 
of photons in UHECR flux are 2\% at $10^{19}\,$eV, 
5\% at $2\times 10^{19}\,$eV, and 30\% at $4\times 10^{19}\,$eV, respectively.
These limits constrain the flux from top-down models to be a sub-dominant 
component of the UHECR flux even above the GZK energy, while they
start to approach at low energies, $E=10^{19}\,$eV,  
the upper range of photons predicted~\cite{dima} by the GZK reaction.
Searches for photons from SHDM at low energies should be therefore 
combined in the future with searches for the anisotropy pattern
connnected to the position of the Sun in the galactic halo.

\begin{figure}
\begin{center}
\epsfig{file=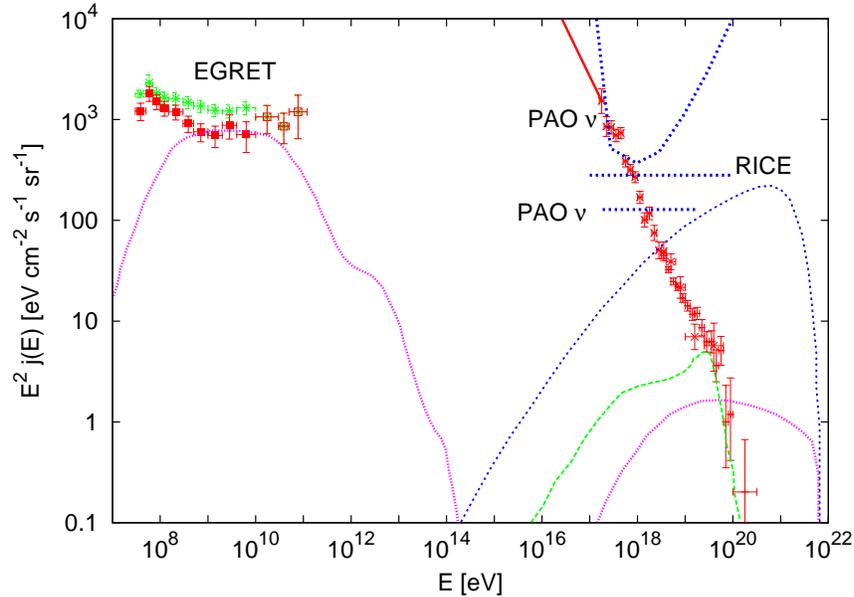,height=11.5cm,angle=270}
\end{center}
\caption{
Proton (green), photon (magenta) and neutrino (blue) fluxes in a TD model 
with $M_X=2\times 10^{13}$~GeV, evolution $\dot n_X \propto t^{-3}$ and 
continuous distribution of sources (adopted from Ref.~[42]) together with 
two determinations of the MeV--GeV diffuse photon background from EGRET data, 
CR data, and the new neutrino limit from PAO [49], (differential top, integral
limit below for an assumed $1/E^2$ neutrino spectrum). \hfill
\label{td}}
\end{figure}

Topological defect models are additionally constrained by the diffuse 
$\gamma$-ray
background and by limits on UHE neutrinos. While it is anticipated that
Fermi/GLAST will resolve part of the diffuse $\gamma$-ray background observed
by EGRET and thus the resulting limits on TD models will become soon more 
stringent, at present there is not yet a change compared to older 
analyses~\cite{ss}. On the other hand, the PAO collaboration derived a 
new limit on the UHE neutrino flux from the non-observation of 
$\nu_\tau$ events~\cite{Abraham:2007rj} that improves the old RICE
bound. This limit constrains $E^2j_\nu(E)$,
i.e.\ the energy contained in neutrinos per energy decade,
by a factor $\sim 10$ stronger than the corresponding EGRET limit  for
$E^2j_\gamma(E)$ the injection of electromagnetic radiation.  
At present the $\nu_\tau$ limit is comparable to the EGRET bound,
cf. Fig.~\ref{td}.

{\em 3. Galactic anisotropy:\/} 
The UHECR flux from SHDM should show a galactic
anisotropy~\cite{Du98}, because the Sun is not in the center of the
Galaxy. The degree of this anisotropy depends for decays only mildly 
on how strong 
the CDM is concentrated near the galactic center. Experiments in the southern 
hemisphere as the old Australian SUGAR experiment~\cite{SUGAR} and the
PAO do see the Galactic center and are thus most sensitive to a 
possible anisotropy of arrival directions of UHECR from SHDM.  
The SHDM hypothesis was found to be compatible with the SUGAR data at 
the $\sim 5$--20\% level\,\cite{Kachelriess:2003rv}. 
The predictions for the SHDM anisotropies
were up-dated\,\cite{up} recently, but not yet used for a limit by the 
PAO collaboration.

Although at low energies, say at $10^{18}$\,eV,
the UHECR flux is strongly dominated by CRs from astrophysical sources,
the search for anisotropies in this energy range may be 
promising\,\cite{Aloisio:2007bh}. The
CR flux from astrophysical sources at  these energies is, apart
from a 0.6\% anisotropy induced by the (cosmological) 
Compton-Getting effect\,\cite{Kachelriess:2006aq}, isotropic and thus 
even the small anisotropy in the few percent level from SHDM decays
may be a detectable signature. Already the 
2005--2006 data\,\cite{Armengaud:2007hc} 
from the PAO presented at the ICRC '07 limit at the 95\% C.L. the amplitude
of the first Rayleigh harmonics to less than 1.7\%  at 1\,EeV and to
3.2\% at 3\,EeV. The corresponding prediction for a SHDM model with
parameters as in Fig.~\ref{ABK}
that respects both photon bounds and fits the observed spectrum are
in the 2\% range\,\cite{Aloisio:2007bh}.
Note however that, since the energy estimate for proton and a photon primaries
differ, the two numbers are not directly comparable and a proper
analysis is required\,\cite{remark}.

\section{Conclusions}

An unavoidable signature of top-down models is the chemical composition
of the resulting ultrahigh energy radiation, because hadronization results
in large neutrino and photon fluxes. The former is an especially prominent
signature for topological defects, while the high photon/nucleon ratio at 
generation is reduced by strong absorption of UHE photons. The resulting 
electromagnetic cascades transfer this energy to the diffuse $\gamma$-ray
background and its future observation by Fermi/GLAST will tighten constraints 
for TD models.

Superheavy dark matter is at present most stringently constrained by
the non-observation of photons, while from the isotropy of the CR flux 
observed by PAO at low energies ($\gsim 10^{18}$\,eV) a competitive limit
may be derived. A combination of both search methods seems to be 
the most powerful way to increase the sensitivity of UHECR experiments
to lower fluxes from SHDM.

\section*{Acknowledgements}
%
I am grateful to all my co-authors for fruitful collaborations and 
many discussions, and to Eric Armengaud, Jose Blanco-Pillado, Markus
Risse and Ken Olum for helpful comments.


\end{document}